# Fingerprint Spectroscopic SRS Imaging of Single Living Cells and Whole Brain by Ultrafast Tuning and Spatial-Spectral Learning

Haonan Lin[1,3], Hyeon Jeong Lee[2,3,5], Nathan Tague[1], Jean-Baptiste Lugagne[1], Cheng Zong[2,3], Fengyuan Deng[2,3], Wilson Wong[1,4], Mary J. Dunlop[1,4] and Ji-Xin Cheng*[,1,2,3]

[1]Department of Biomedical Engineering, [2]Department of Electrical & Computer Engineering,

[3]Photonics Center, [4]Biological Design Center,

Boston University, Boston, MA 02215, USA

[5]College of Biomedical Engineering and Instrument Sciences, Zhejiang University, Hangzhou, China

*corresponding author

Contact information:
Ji-Xin Cheng
email: jxcheng@bu.edu
phone: +1 617-353-1276

**ABSTRACT**

Label-free vibrational imaging by stimulated Raman scattering (SRS) provides unprecedented insight into real-time chemical distributions in living systems. Specifically, SRS in the fingerprint region (400–1800 cm$^{-1}$) can resolve multiple chemicals in a complex bio-environment using specific and well-separated Raman signatures. Yet, fingerprint SRS imaging with microsecond spectral acquisition has not been achieved due to the small fingerprint Raman cross-sections and the lack of ultrafast acquisition scheme with high spectral resolution and high fidelity. Here, we report a fingerprint spectroscopic SRS platform that acquires a distortion-free SRS spectrum with 10 cm$^{-1}$ spectral resolution in 20 µs using a lab-built ultrafast delay-line tuning system. Meanwhile, we significantly improve the signal-to-noise ratio by employing a spatial-spectral residual learning network, reaching comparable quality to images taken with two orders of magnitude longer pixel dwell times. Collectively, our system achieves reliable fingerprint spectroscopic SRS with microsecond spectral acquisition speed, enabling imaging and tracking of multiple biomolecules in samples ranging from a live single microbe to a tissue slice, which was not previously possible with SRS imaging in the highly congested carbon-hydrogen region. To show the broad utility of the approach, we have demonstrated high-speed compositional imaging of lipid metabolism in living pancreatic cancer Mia PaCa-2 cells. We then performed high-resolution mapping of cholesterol, fatty acid, and protein in the mouse whole brain. Finally, we mapped the production of two biofuels in microbial samples by harnessing the superior spectral and temporal resolutions of our system.

**Introduction**

Stimulated Raman scattering (SRS) microscopy is a high-speed vibrational imaging modality that produces chemical maps in dynamic living systems based on intrinsic molecular vibrations[1–5]. Such capability allows direct visualization of complex biological processes without perturbation, enabling a plethora of biomedical applications, such as tracking voltage spiking during neuron firing[6], identifying the cancer margin of fresh, unprocessed tissues[7], and discovering biomarkers and therapeutic targets of aggressive cancers[8,9]. When evaluating an SRS system, speed, spectral bandwidth and signal-to-noise ratio (SNR) are the three major aspects, which characterize the temporal resolution, chemical specificity, and reliability. Utilizing narrowband pump and Stokes lasers, single-color SRS has reached the speed of up to video-rate[10]. Meanwhile, spectroscopic SRS has been developed to acquire a Raman spectrum at each pixel, enabling the simultaneous study of chemicals with overlapping Raman bands in complex biological samples. Spectroscopic SRS has been achieved in 3 ways, namely spectral scanning of a narrowband laser[11,12], parallel detection of a complete spectrum by a detector array[13], and temporal scanning of linearly chirped pulses[14]. To date, spectroscopic SRS can record a spectrum within a ~200 cm$^{-1}$ window at the microsecond level[14]. Despite major advances in instrumentation that push the speed and the spectral bandwidth, most SRS applications are focused on the carbon-hydrogen (C-H) stretching region (2800 – 3100 cm$^{-1}$) where strong Raman bands reside. However, the highly crowded SRS signals in the C-H region severely limit the chemical specificity of SRS in a complex biological environment. Fingerprint SRS can significantly enhance the chemical specificity by providing a specific and well-separated Raman spectrum for each biochemical component. However, the much weaker Raman cross-section in the fingerprint region results in a decrease in the signal level under the same imaging conditions as the C-H region. To maintain high imaging speed, one can increase the laser power within the damage threshold to prevent the signal from being overwhelmed by noise. Thus, the proposed spectral acquisition scheme needs to have high laser power efficiency. In addition, the fingerprint Raman peaks for different biochemical are narrow and close to each other, which imposes the requirement for high spectral resolution. Therefore, to acquire high-fidelity fingerprint SRS spectra at the microsecond level, a high-speed spectral acquisition scheme that can achieve both high power efficiency and high spectral resolution is required. Among the spectral acquisition schemes, spectral focusing is the most power-efficient since all energy of the femtosecond pulses is used. However, existing high-speed spectral focusing scheme[14] by an edge-reflected resonant mirror only has 2 ps delay range, which limits the degree of chirping and leads to 28 $cm^{-1}$ spectral resolution that is insufficient for the fingerprint region. Up to date, a scheme that can acquire fingerprint SRS spectra at the microsecond level with a spectral resolution below 10 $cm^{-1}$ has not been reported, which inhibits spectroscopic fingerprint SRS from being a reliable tool for broad applications.

Due to the physical limits, advances of instrumentation alone are not enough to achieve reliable high-speed fingerprint spectroscopic SRS imaging. The physical limits lead to the trade-offs between speed, spectral bandwidth and SNR, which can be conveniently expressed as a 3D hyperplane design space (**Fig. 1a**). Various computational methods have been proposed to extend the design space. Matrix completion[15,16] and compressed sensing[17,18] methods have been used to sub-sample images to increase speed while avoiding information loss. Denoising algorithms with models[19,20] on object structures have also been proposed to recover the SNR of microscopic images with low light exposure or low pixel dwell times. Most computational methods depend on the formulation of forward models to describe the underlying imaging process, such as the modulation of measurements by a mask, the blurring of the image by the optical point-spread function, the thermal and electronic noise of photodetector and the laser shot noise. However, formulating a forward model requires tedious system calibration, and certain simplifications are necessary for the sake of computational tractability. A method that can bypass model design and directly learn features of the image to formulate mappings from raw experimental data to reliable results should outperform the abovementioned unsupervised computational methods. Deep learning[21] is such a method and offers an appealing approach to this problem. Given training data of input/output image pairs, a deep neural network learns the nonlinear mappings that find optimal approximate solutions to a variety of complicated inverse problems that are challenging to address using conventional analytical methods. Deep learning has been applied to a broad range of vibrational imaging applications, such as image restoration of single-color SRS images in the C-H region under low light exposure[22] and automated detection of the tumor margin from fresh tissue[23]. However, little has been done to utilize deep learning for processing spectroscopic SRS images which are 3D image stacks with unique spectral features that are different from volumetric data. Directly applying 3D convolution neural network (CNN) on spectroscopic data fails to consider the different physical correlations of spatial and spectral domain, which may introduce artifacts and degrades recovery quality. In addition, training a deep network with 3D CNN filters requires very high computation cost and is difficult to train. Thus, a novel convolution filter design for hendeling spectroscopic image datatype is much needed to facilitate deep learning as a practical tool to push the physical limits and deliver a fingerprint spectroscopic SRS system with a much higher speed and spectral fidelity level.

Herein, we demonstrate high-fidelity fingerprint spectroscopic SRS imaging scheme with microsecond spectral acquisition. Such capability was enabled by the combined innovations of ultrafast delay-line tuning instrumentation and image restoration by a spatial-spectral residual net (SS-ResNet). First, we realized a high-speed and high-spectral resolution spectral focusing scheme by incorporating a 55-kHz polygon scanner and a Littrow-configured reflective grating as the delay-line scanner (**Fig. 1b**). We note that the polygon Fourier-domain delay line scanning was implemented for FT-CARS spectroscopy[24]. In comparison, our setup has two key innovations that significantly improve the versatility and reliability. Firstly, our scheme maintains a very high linearity between the data sampling from the trigger and corresponding Raman shifts, which makes the recorded spectrum free of distortion. Secondly, by rotating the blazed grating to change the angle between the laser-scanned line and the blazed line of the grating (angle θ in **Fig. 1b**), the maximum delay is tunable, which allows perfect match between the delay range and an arbitrary degree of pulse chirping. To compensate for the signal level decrease due to higher imaging speed and extensive chirping, we further applied a deep neural network to reconstruct SRS spectroscopic images from high-speed, low-SNR raw images in the fingerprint region. Inspired by the pseudo-3D network for video processing[25], we consider the different physical correlations between spatial and spectral domains and replaced the widely-used $3 \times 3 \times 3$ 3D CNN filter by two parallel filters: A $1 \times 3 \times 3$ convolution filter on the spatial domain to capture spatial correlations and a $3 \times 1 \times 1$ convolution filter on the spectral domain to maintain spectral continuity between adjacent frames. Consequently, SS-ResNet reduces the training model size, which facilitates the training of a deep network. More importantly, spatial-spectral crosstalk distortions can be avoided, which improves the reconstruction accuracy compared to 3D CNN. Finally, we deployed a pixel-wise least absolute shrinkage and selection operator (LASSO) regression[26] algorithm to decompose the recovered spectroscopic image into maps of different biomolecules. Compared to the previous linear unmixing method by least-square fitting[27], our pixel-wise LASSO unmixing can effectively suppress the crosstalk between different chemical maps by incorporating the prior knowledge that at each location only a few components have dominant contributions. Our results show that this workflow is an effective approach for mapping multiple chemical species with high-speed and high-fidelity using the fingerprint region, allowing real-time quantitative mapping of chemical compositions in living biological systems. To demonstrate the capability of our scheme, we performed real-time imaging of lipid species, including cholesterol and unsaturated fatty acids, in living cancer cells. Furthermore, our high-speed imaging technique allowed large-area mapping of biomolecules in the mouse whole brain, revealing distinctive distributions of fatty acid and cholesterol in nerve bundles and populations of cholesterol-rich cells in certain brain regions. Finally, we demonstrated the capability to differentiate multiple biomolecules by imaging biofuel production by engineered microbes. These results and applications collectively show high-speed, high-fidelity fingerprint spectroscopic SRS imaging and its potential in addressing a plethora of significant biomedical and bioengineering problems.

**Results**

**Spectroscopic SRS by polygon scanner and Littrow-configured grating**

The concept of ultrafast tuning is illustrated in **Fig. 1b**. Briefly, two femtosecond lasers (pump and Stokes) are linearly chirped by high dispersion medium to temporally separate different frequency components. The Stokes beam is sent to a 55-kHz polygon scanner and subsequently scanned to a blazed grating at Littrow configuration, which acts as a wedge to introduces a continuous-changing path difference between the pump and the retroreflected Stokes beam. Consequently, an SRS spectrum can be acquired within 20 µs. A detailed description of the optical setup is provided in **Methods** and depicted in **Supplementary Fig. 1a**. Importantly, by rotating the blazed grating (**Fig. 1b**) to change the angle between the laser scanning line and the grating blazed line, which shortens the effective delay range. Consequently, the delay tuning range is adjustable from 0 ~ 20 ps (the maximum range is determined by the grating blazed angle), which allows for extensive chirping of the lasers for dramatically improved spectral resolution. In the current system, the longer delay range enables the use of 90-cm SF57 glass rods after the beam combiner, resulting in a spectral resolution of 10 cm$^{-1}$ in the fingerprint region (**Supplementary Fig. 2a-b**). Such spectral resolution is essential for resolving multiple chemicals in a fingerprint window. In addition, given the linear speed of the polygon scanner, the acquired raw Raman spectrum is free of spectral channel distortion. For evaluation, we measured the spectral profiles of five chemicals and compared them with spontaneous Raman spectroscopy (**Supplementary Fig. 2c-d**). Eleven significant peaks were used to map the sampling points from triggering to the Raman shifts from Raman spectroscopy, showing high linearity with $R^2 = 0.9997$ (**Supplementary Fig. 2e**). The sensitivity was quantified by acquiring SRS spectra from dimethyl sulfoxide (DMSO) diluted with DI water (**Supplementary Fig. 2f**). Besides the background due to cross-phase modulation, the DMSO solutions contributed to a significant peak at 2913 cm$^{-1}$. At concentrations as low as 0.125% v/v, the DMSO peak was still separable from the background, suggesting a high sensitivity in the C-H region. However, in the fingerprint region, excessive averaging is necessary to obtain an SRS spectrum with high SNR (**Supplementary Fig. 2g-h**).

**SNR recovery via spectra-spatial residual net and chemical mapping by pixel-wise LASSO unmixing**

To extract information from the high-speed yet noisy spectroscopic images, we apply a two-step processing approach that involves SNR recovery and chemical mapping. To recover the SNR, we deployed a deep neural network, acting as a supervised denoiser, to recover the SNR of high-speed fingerprint SRS images (**Fig. 1c**). We first generated pairs of spectroscopic SRS images as the training set, with high-speed, low SNR images as the raw acquisition and a low-speed, high-SNR image (through averaging of multiple raw acquisitions) as the ground truth. As shown in **Supplementary Fig. 3a**, considering the large size of each spectroscopic image and the experimental difficulty of generating a large number of training images, the network is based on the U-net[28] encoder-decoder structure. The up-sampling and skip-connection layers in the network improves the resolution of learned features and thus requires less training samples. We do not directly apply 3D CNN kernels to spectroscopic images. Instead, we replace the $3 \times 3 \times 3$ convolution kernels by two parallel CNN kernels, including a $1 \times 3 \times 3$ spatial convolution filter and a $3 \times 1 \times 1$ spectral convolution filter to maintain spectral continuity between adjacent frames (**Supplementary Fig. 3a**). Finally, residual learning scheme[29] is applied to facilitate the training of a deep network.

After SNR recovery, the spectroscopic image stack was linearly decomposed into chemical maps (**Fig. 1d**) to facilitate downstream visualization and analysis. Based on the observation that at each spatial location, only a few chemical components have dominant contributions, we used pixel-wise LASSO regression[26], which incorporates individual $l_1$- norm sparsity regularization to the concentrations at each pixel. The level of regularization can be fine-tuned such that the output can suppress cross-talks between different channels while avoiding artifacts. We applied the approach to unprecedented imaging conditions reaching high speed, high SNR and high chemical specificity in the fingerprint region for a vast variety of biological samples, including living cancer cells, mouse whole brains, and single bacteria, with a focus on chemicals that are difficult to study in the C-H region.

**High-speed spectroscopic fingerprint SRS imaging of lipid metabolism in Mia PaCa-2 cells**

Lipid metabolism is a cellular process involving spatiotemporal dynamics of fatty acid and cholesterol. The distributions of different lipid species in the cell are tightly regulated to ensure proper cellular activities and function. Abnormal lipid metabolism is related to many human diseases including aggressive cancer[8,9]. Thus, quantitative imaging of lipids in living systems is of great interest. Unlike fluorescence imaging by lipophilic dyes, Raman spectroscopy provides high chemical specificity to differentiate lipid species, such as cholesterol and various fatty acids. With enhanced signal levels, SRS is capable of quantitative imaging of specific lipid species. For example, cholesterol imaging has been demonstrated in cholesterol-rich samples such as the atherosclerotic artery[30] and lysosome-related organelles in *C. elegans*[31] by focusing on the sterol C=C stretching band at 1669 $cm^{-1}$. However, due to the limited signal levels in the fingerprint region, except in the abovementioned cases of excessive accumulation, it remains challenging to study cholesterol in living cells or large-area tissues.

To demonstrate that our system can achieve real-time lipid tracking in living cells, we imaged Mia PaCa-2 cells within the 1550 - 1750 $cm^{-1}$ fingerprint vibrational window. For training, we first acquired a dataset consisting of pairs of raw and ground truth images of Mia PaCa-2 cells. We used fixed Mia PaCa-2 cells to ensure that the ground truth images formulated by excessive averaging do not suffer from motion artifacts. Each raw spectroscopic image stack covering a ~200 $cm^{-1}$ spectral window with $200 \times 200 \, \mu m^2$ field-of-view (FOV) was acquired within 1.8 seconds. The ground truth image was generated by averaging 100 raw images of the same FOV, resulting in a ~10-fold SNR enhancement. After training, the performance of SNR recovery was validated using a set of previously unseen images. Comparing the raw, network recovered and ground truth images of the same FOV at 1650 $cm^{-1}$ (**Fig. 2a-c**), we demonstrated that the SS-ResNet recovery allows reconstruction of the raw spectroscopic image stack, reaching comparable image quality to the ground truth images. To justify the performance of SS-ResNet recovery, the same validation dataset was processed by block-matching 4D filtering (BM4D), a state-of-the-art unsupervised 3D image denoising algorithm. Also, to compare the performance of spatial-spectral convolution, a network with identical structure except for replacing the SS-Conv with 3D CNN was trained and tested on the same dataset. The results (**Supplementary Figure 4a-e**) suggest that both networks outperformed BM4D significantly. Meanwhile, the SS-ResNet is better than 3D CNN by maintaining more detailed structures without introducing artifacts. We further quantified the observations by calculating the normalized root mean square error (NRMSE) and structural similarity (SSIM) index[32] for the raw vs. ground truth, BM4D vs. ground truth 3D CNN vs. ground truth and SS-ResNet vs. ground truth (**Supplementary Figure 4f-g**). Both measurements suggest significant improvement of the image quality using SS-ResNet. To test whether the network recovery facilitates downstream spectral analysis, we selected a small region of interest from the validation set (**Fig. 2d**) and performed pixel-wise LASSO unmixing on raw, SS-ResNet and ground truth image stacks using three SRS spectral profiles generated from Bovine serum albumin (BSA), triglyceride and cholesterol (**Fig. 2e**). These spectral profiles represent 3 major chemical bonds, including the Amide I band at 1650 cm$^{-1}$ from proteins, the acyl C=C band from lipid acyl chains at 1650 cm$^{-1}$, and the sterol C=C band from cholesterol a 1669 cm$^{-1}$. The outputs from the network and the ground truth show similar spatial distributions and concentrations for all 3 components. In contrast, the results from the raw data failed to provide insights into the distributions of chemical species and were difficult to distinguish from the background noise (**Fig. 2f**). To quantify the quality of chemical maps after network recovery, we calculated the SSIM index for all the three chemical channels (**Fig. 2g**). The SSIM indices increased considerably after recovery, which proved that our approach did not introduce artifacts and provided reliable results on the subsequent chemical analysis.

To apply this high-speed, high-sensitivity method to the real-time mapping of lipid in living cells we imaged living Mia PaCa-2 cells and recovered high-resolution images from the raw images taken at high speed by applying the same SS-ResNet trained on fixed cells. In living Mia PaCa-2 cells, lipid droplets are shown to be highly dynamic[33]. Live-cell imaging at the speed of 1.8 seconds per stack was performed on Mia PaCa-2 cells to capture lipid droplet dynamics. Indeed, we observed severe motion artifacts in the 100-averaged image from the live-cell data (**Fig. 2h**). SS-ResNet recovered images from a single frame showed clear circular-shaped droplets within the cells, highlighting the importance of temporal resolution during live-cell imaging. The chemical maps of cholesterol and fatty acid (**Fig. 2i-j**) further confirmed that motion artifacts affect the fidelity of the subsequent spectral analysis. After recovery, clear lipid dynamics can be visualized at $1650\ cm^{-1}$ (**Supplementary Video 1-2**), and real-time chemical mapping of protein, cholesterol and fatty acid can be achieved (**Supplementary Video 3-5**).

We further asked whether this method could be used to track changes in cholesterol amount and distribution. To that end, we imaged two sets of living Mia PaCa-2 cells, a control set and a set treated with HP$\beta$CD, which extracts cholesterol from the cell membrane[34]. Compared with the control group, the cholesterol concentration in the cell membrane decreased significantly after HP$\beta$CD treatment, whereas the fatty acid concentration was maintained at the same level (**Fig. 2k-n**). The remaining cholesterol after HP$\beta$CD treatment mainly distributed within the lipid droplets. By calculating the single-cell ratio between cholesterol and fatty acid concentrations for ~1000 cells from the control and the HP$\beta$CD-treated groups, we confirmed significant reductions in cellular cholesterol after the treatment (**Fig. 2o**). These data show that deep-learning high-speed fingerprint SRS imaging enables high-fidelity, real-time chemical mappings of chemical bonds in single living cells and facilitates the tracking of metabolite dynamics at subcellular levels.

**Chemical mapping of a whole mouse brain slice by high-throughput spectroscopic fingerprint SRS imaging**

Brain tissue is comprised of many cell types, and biomolecules in the tissue are highly heterogeneous among different brain areas. Chemical mapping of the whole brain is essential for studying the functionality of molecules in the brain. Previous label-free metabolic studies of mouse whole brain slices were mainly based on multi-color SRS imaging in the C-H window, providing only protein and lipid information[7,35]. For the sake of maintaining sample conditions during the experiment, the total acquisition time of a mouse whole brain slice is usually several hours. Therefore, it remains challenging to perform spectroscopic SRS imaging in the fingerprint region to generate chemical maps of other biomolecules.

Following the procedures in **Fig. 1c**, we first took a training dataset in different brain regions, including the lateral hypothalamus (LH), caudate putamen (CPu), cortex (CTX), habenula (HB), medial habenula (MH), ventral lateral nucleus (VL), hippocampus (HC), dentate gyrus (DG) and corpus callosum (CC). Each raw image was taken at a speed of 3.8 seconds per spectroscopic image stack with a $200 \times 200 \ \mu m^2$ FOV and the high-SNR ground truth GT image was acquired by averaging the raw measurements of the same FOV 100 times (**Supplementary Fig. 5**). After training, a validation set was used to test the ability to recover SNR using SS-ResNet. After recovery, the SNR of the raw image improved significantly while the subcellular details are preserved, reaching comparable image quality to the ground truth image (**Fig. 3a-c**). To quantify the reconstruction quality, we measured the NRMSE and SSIM for the raw vs. ground truth and SS-ResNet vs. ground truth for 15 different validation images (**Fig. 3d**). Taking advantage of the high imaging speed of our system and the ability to recover high SNR by SS-ResNet, we performed fingerprint SRS spectroscopic imaging of a mouse whole-brain slice. Acquisition of the whole brain slice over a ~200 cm$^{-1}$ spectral window in the fingerprint region was finished within 3.5 h, which is comparable to the acquisition time of multi-color SRS imaging in the C-H region focusing on a few Raman shifts[7]. The comparison between the raw image and the network recovered the image of the whole brain tissue at 1650 cm$^{-1}$ demonstrates that morphologies of single cells and nerve bundles within the brain can be clearly distinguished after recovery (**Supplementary Fig. 6**).

We further applied pixel-wise LASSO spectral analysis of the SS-ResNet recovered image stack to produce chemical maps of the amide I group (blue, for protein), acyl C=C (red, for unsaturated fatty acid) and sterol C=C (green, for cholesterol) for the whole brain slice (**Fig. 3e-g**). The composite image of the 3 components shows significant heterogeneity among different cells and brain structures (**Fig. 3h**), reflecting a relative abundance of protein, fatty acid and cholesterol. To further characterize the distribution of the biomolecules, we focused on several brain regions and features (**Fig. 3i**). Overall, the soma of mature neurons shows relatively lower concentrations of all three components compared to the surrounding tissue. Surprisingly, we found abundant cholesterol-rich cells present near neurons in the LH and basal amygdaloid (BM) regions, which may represent different metabolic activities in this population of cells. We also observed that nerve bundles in the ventral posterior nucleus (VP) and CPu are comprised of different ratios of cholesterol and fatty acid. Interestingly, there are a few rare cells that contain high cholesterol concentrations in the DG region (Circled regions in **Fig. 3i**). As DG is one of the regions containing neural stem/progenitor cells, we suspect that these cholesterol-rich cells may reflect cells undergoing hippocampal neurogenesis. In summary, the large-area imaging in the fingerprint region is a viable tool for the label-free study of the cellular cholesterol content, which could be used to address many important biomedical questions about the relationship between cholesterol metabolic activity and a variety of brain diseases and disorders, including neurodegenerating disorders and brain tumors.

**High-throughput spectroscopic fingerprint SRS imaging of *E. coli* biofuel production**

Limonene and pinene are biofuel precursors that can be produced biosynthetically in microbes such as *Escherichia coli* (*E. coli*) using strains that have been engineered to produce the enzymes necessary to synthesize these chemicals[36–38]. Currently, quantitation of biochemical production levels mainly relies on gas chromatography-mass spectrometry (GC-MS), which suffers from low throughput and requires extraction steps that destroy the sample. Strain engineering and optimization typically involve the construction of many variants, followed by screening, in a lengthy iterative process. The limited throughput of GC-MS approaches hinders efficient optimization of design variables for biochemical synthesis. In addition, GC-MS only provides quantification of population-level production, ignoring the potential for genetic or phenotypic variation among cells[39,40]. Thus, a high-throughput quantification method that provides direct measurement of biofuel concentrations has the potential to improve the design, build, and test cycle necessary for improving production strains. SRS is a promising approach to fulfill this requirement by detecting intrinsic vibrational signatures from the biofuels that are linearly related to the concentrations. Yet, due to the overwhelming SRS contributions from endogenous proteins and lipids, quantitative imaging of the production levels for certain biofuels (i.e., limonene, pinene) in the crowded C-H region has been challenging. High-throughput SRS imaging in the fingerprint region is expected to address this challenge by providing specific and well-separated Raman spectra for the biofuels.

Our platform provides a viable approach towards high-throughput quantitative chemical imaging of chemical compounds produced biosynthetically by bacteria. **Fig. 4a** depicts the ~1650 cm$^{-1}$ fingerprint Raman window, in which unsaturated fatty acid contributes to the peak at 1655 cm$^{-1}$, limonene has two peaks at 1645 cm$^{-1}$ and 1678 cm$^{-1}$, while $\alpha$-pinene contains a peak at 1660 cm$^{-1}$. The peaks all originate from C=C bonds but differ from each other due to the specific structures of each chemical. Additionally, the amide 1 group from protein contributes a broad Raman band around 1650 cm$^{-1}$, serving as the contrast for the cell body. Using the same strategy as in the previous two applications, we acquired training and testing sets from both wild type cells and biofuel production strains, which consisted of pairs of high-speed, low SNR and low-speed, high-SNR images through 50 averages. After training, SS-ResNet was applied to a validation set to test the recovery performance. Examples of validation images at 1650 cm$^{-1}$, including the raw image, SS-ResNet recovery and ground truth are shown in **Fig. 4b-d**. Quantitation of the reconstruction quality is depicted in **Fig. 4e**, suggesting that it is possible to denoise images while maintaining high-quality spatial localization data. Finally, we performed high-speed imaging and SS-ResNet recovery on images of a wild type strain (*E. coli* BW25113), which does not produce biofuel. We compared this to limonene production[37] and pinene production[38] strains of *E. coli*. Based on the spectral profiles from pure chemicals, pixel-wise LASSO spectral analysis decomposed the network-recovered spectroscopic images of the strains into the maps of the three chemicals.

The chemical maps indicated that the wild type strain (**Fig. 4e**) only had significant signals from the protein in the cell bodies. Whereas the limonene (**Fig. 4f**) and pinene (**Fig. 4g**) producing strains had protein signals and a substantial increase in the corresponding concentrations of intracellularly aggregated chemicals. We did not include fatty acid and cholesterol in the analysis due to the negligible contributions. Using our scheme, the acquisition time of fingerprint SRS imaging was 8 seconds for a $50 \times 50\ \mu m^2$ FOV covering hundreds of *E. coli* cells, offering excellent potential for high-throughput screening to optimize the design variables of biofuel production pathways.

**Discussion**

We have demonstrated a novel implementation of a vibrational SRS spectroscopic imaging scheme that reaches an ultrafast acquisition speed of 20 µs per SRS spectrum. Compared with our previous spectral focusing implementations using a 12-kHz resonant scanner[14], the polygon scanning system not only improves the speed 5-fold but also achieves high spectral linearity that increases reliability. More importantly, the delay range of our previous configuration was fixed to ~2 ps, which allowed only a moderate degree of chirping and thus reached a spectral resolution of 28 cm$^{-1}$. In comparison, our tunable delay range enables a much higher degree of chirping, reaching a spectral resolution within 10 cm$^{-1}$ in the fingerprint region. Currently, the spectral coverage is ~ 200 cm$^{-1}$ due to the spectral bandwidth of the laser sources. However, since the delay range is freely tunable, if combined with broadband lasers by fiber amplification[41] or supercontinuum laser sources, the scheme can potentially be used to obtain the entire fingerprint SRS spectrum within 20 µs. In addition, the high-speed tunable delay scanning scheme can also be applied to a broad range of modalities requiring a long delay scan, such as transient absorption spectroscopy and impulsive SRS imaging.

In this work, we trained a spatial-spectral residual net as a supervised denoiser that outperformed conventional unsupervised image restoration algorithms. The encoder-decoder structure alleviates the requirement for training data size, which is of great importance for biomedical imaging given the high cost associated with acquiring training data. Here, fewer than 20 spectroscopic images (200 × 200 × 128 size for each image) were used as the training set for each application. We note that a similar U-net residual learning design has been successfully applied to denoise volumetric fluorescent images[42]. In this work, we demonstrate that for spectroscopic images, spatial-spectral convolution kernels can outperform 3D CNN in terms of accuracy. Our supervised denoiser can significantly increase the reliability of the subsequent chemical content analysis. Besides, for the task of denoising spectroscopic image stacks, due to the universal properties of noise under the same imaging conditions, a trained network can be quickly tweaked to denoise other samples by transfer learning. As shown in **Supplementary Fig. 7**, we applied a network pre-trained on Mia PaCa-2 cells to recover prostate tissue images taken under the same imaging conditions. Direct application achieved high SNR levels but sacrificed spatial resolution due to the differences between spatial features for the two datasets. By feeding in training data of the new samples, the network required less than half of the training epochs to converge and output high-resolution, high-SNR images, making it convenient to apply to different applications.

In conclusion, the combined use of ultrafast delay-line scanning and deep learning has enabled reliable fingerprint SRS imaging with microsecond spectral acquisition. Using the network, an effective sensitivity enhancement by an order of magnitude is achieved. The improved spectral resolution by the instrumentation and the improved SNR by the network are essential for fast and reliable SRS imaging in the fingerprint region. With such advances, we have demonstrated imaging of various biomolecules that are difficult to identify in the high-wavenumber C-H window. This technique has broad applications, as demonstrated in this study: from monitoring biofuel production levels in engineered bacteria to the metabolic study of cancer cells, up to large-area whole brain tissue imaging. Collectively, our approach opens the door to a plethora of biomedical applications from tracking dynamics and interactions of metabolites in a single cell to the high-throughput compositional mapping of unprocessed human tissue.

## Methods

### Ultrafast tuning spectroscopic SRS microscopy

The ultrafast tuning SRS microscopy is illustrated in **Supplementary Fig. 1a**. A dual-output 80-MHz femtosecond pulsed laser (InSight DeepSee+, Spectra-Physics) was used to deliver synchronized pump and Stokes lasers. The 120-fs tunable laser (680 – 1300 nm) was used as the pump beam while the 200-fs output fixed at 1040 nm served as the Stokes beam. The Stokes beam was modulated by an acousto-optical modulator (AOM, 1205-C, Isomet) at 2.4 MHz for heterodyne detection. The Stokes beam was then directed to a polygon scanner (Lincoln SA24, Cambridge Technology), which scanned the laser onto a blazed grating positioned at Littrow configuration. The grating acted as a reflective wedge to reflect the Stokes beam along the same optical path. Each scan by the polygon scanner thus introduces a continuous increase of light path for a few millimeters, resulting in a series of continuous temporal delays between the pump and the retroreflected Stokes beam. The maximum delay range (denoted as L) of the system is determined by the length of the scan line and the blazed angle of the grating. By rotating the grating to change the angle between the scan line and the blazed line (denoted as θ), the effective delay is reduced to $Lsin\theta$. The beams were combined by a dichroic mirror (DM, Chroma) and were both broadened to picosecond by high dispersion glass rods (SF57) with 90 cm in total length. The chirped beams were sent collinearly to an upright microscope and a 2-D galvo scanner set (GVS102, Thorlabs) was used for scanning images. A 60X, 1.2 NA water immersion objective (UPLSASP 60XW, Olympus) was used to focus the light onto the sample followed by forward collection by an oil immersion condenser. After filtering the Stokes beam after interacting with the sample, a photodiode (S3994-01, Hamamatsu) with a custom-built resonant circuit was used to collect signals. The SRS signal was extracted by a lock-in amplifier (UHFLI, Zurich Instrument) and was digitized by a high-speed digitizer (ATS 460, AlazarTech). A custom-written Matlab (MathWorks) code was used to synchronize the scanning of spectrum with the polygon scanner and the scanning of the galvo mirrors to generate **Error! Bookmark not defined.**spectroscopic image stack in a $\lambda - XY$ manner.

**Deep-learning network structure, training and error quantification**

As depicted in **Supplementary Fig. 3a**, the network for SNR recovery of the spectroscopic image is based on U-net, which consists of an encoder-decoder architecture. At each layer, a $1 \times 1 \times 1$ convolution layer is first used to increase the feature dimensions, followed by a total of 6 spatial-spectral convolution (SS-Conv, **Supplementary Fig. 3b**) layers. A max-pooling layer is applied at the end of each layer to reduce the dimensions. In the decoder phase, each layer first up samples the feature map and then concatenates it with the corresponding feature maps in the encoder phase. The same 6 SS-Conv layers are used at each layer. At the final stage, a $1 \times 1 \times 1$ convolution layer with linear activation was used to map the feature maps into the prediction of pixel values of the high-SNR image. In addition, the prediction layer was added with the input layer such that the prediction value was the residual[29] with respect to the raw input image, which has been shown to predict higher resolution images. The parameters were learned by minimizing a loss function that averages the mean squared error between the prediction and ground truth. The network was implemented using Keras with Tensorflow as backend and was trained using a graphics processing unit (GPU, RTX 2080 Ti, Nvidia).

To quantify the reconstruction error and compare it with the raw input, we first normalized the ground truth and the predicted image to the same dynamic range by the same method reported in ref. [42]. We then calculated the normalized root mean squared error (NRMSE) and structural similarity index (SSIM; a metric that measures the similarities between two images, higher is better) using the normalized image pairs. Using the same procedure, we calculated NRMSE and SSIM values between the raw input and ground truth in comparison.

**Linear unmixing of spectroscopic images using pixel-wise LASSO**

Assuming the dimensions of the spectroscopic image in $x, y, \lambda$ as $N_x, N_y, N_\lambda$, we first rearrange the 3D spectroscopic image stack as a 2D data matrix ($D \in \mathbb{R}^{N_x N_y \times N_\lambda}$) by arranging the pixels in the raster order. Given the number of pure components as $K$, a bilinear model is used to decompose the data matrix into the multiplication of concentration maps $C \in \mathbb{R}^{N_x N_y \times K}$ and spectral profiles of pure chemicals $S \in \mathbb{R}^{K \times N_\lambda}$:

$$D = CS + E, \qquad (1)$$

where $E$ is the residual term. To simplify the problem, we obtained $S$ by measuring the spectral profiles from pure chemicals. The concentrations can be obtained by minimizing the error term $E$ through the least-squares fitting. However, in practice, least-squares fitting alone generates chemical maps with severe cross-talks in complex biological samples where many biochemicals have overlapping spectral profiles. To improve the performance, we observe that for each spatial pixel, only a few chemical components contribute significantly, which is equivalent to the sparsity of concentrations at each pixel. Thus, we introduced a $l_1$ norm regularization to the original least-squares fitting problem, leading to the following optimization problem which solves for the optimal solution $\hat{C}$:

$$\hat{C} = \arg\min_C \{\frac{1}{2} ||D - CS||^2 + \beta \sum_{i=1}^{N_x N_y} |C_{i,:}|\}, \qquad (1)$$

where $\beta$ is a hyper-parameter controlling the level of the sparsity of the concentration maps and $C_{i,:} \in \mathbb{R}^K$ is the vector containing all the concentration values at a spatial pixel location. For a set of data recorded in the same imaging and digitizing conditions, the value of $\beta$ needs tuning only once. The method, known as the least absolute shrinkage and selection operator (LASSO), has been widely used to solve problems in which the variable is sparse, e.g., compressed sensing[43]. With the use of LASSO unmixing, it is possible to resolve more chemicals in the same window since LASSO effectively suppresses the cross-talks between different channels, leveraging the required concentrations of the chemicals to be identified in a complex living system.

**Mia PaCa-2 cancer cell preparation**

Mia PaCa-2 cancer cells were grown in a monolayer at 37 ˚C in 5% $CO_2$ in RPMI-1640 medium supplemented with 10% fetal bovine serum. To prepare fixed cell samples for training, Mia PaCa-2 cells were cultured on a glass-bottom dish for 1 – 2 days at the humidified chamber and were fixed with 10% neutral buffered formalin for 15 minutes at room temperature. The cells were then washed with and imaged in PBS buffer. For cholesterol depletion in Mia PaCa-2 cells, 500 µM HPβCD was added to the medium and cultured for 24 hr.

**Brain tissue preparation**

The mouse brain slice was prepared from a mouse (Jackson Lab) at age 21 days. PBS was used for perfusion, after which formalin was perfused to fix the brain tissue. Then the brain tissue was frozen sectioned at 150 µm thickness.

***E. coli* biofuel strains**

The *E. coli* strains used in this study are derived from strain JW0451-2 (K-12 BW25113 Δ*acrB*) from the Keio collection[44]. The kanamycin resistance marker gene was removed from the

Keio collection strain. This "wild type" strain was then transformed with plasmids expressing the heterologous pathways for either pinene or limonene production. For pinene production, the chassis strain was transformed with two plasmids, pJBEI-3933 & pJBEI-3085[38], that were gifts from Jay D. Keasling. For limonene production, the chassis strain was transformed with plasmid pJBEI-6409[37], provided by Taek Soon Lee via Addgene (#47048).

Prior to SRS imaging, overnight cultures were inoculated in Luria Bertani (LB) medium with appropriate antibiotics for plasmid maintenance and refreshed the following day in 5mL of M9 minimal media supplemented with 20g/L glucose and appropriate antibiotics. When the cultures reached an $OD_{600}$ (optical density at 600nm) of 0.6, pinene or limonene production was induced by adding IPTG to the culture (500 µM and 25 µM, respectively). The cultures were grown at 37 ˚C for another 18-24 hours. 5-10 minutes before imaging, 5 µL of culture was placed on a 3% agarose pad and pressed between microscope coverslips to immobilize the cells, and then the sample was imaged.

**Author Contributions:** H.L., C.Z., and F.D. designed and implemented the ultrafast tuning system. H.L. wrote the code for SS-ResNet SNR recovery and LASSO spectral unmixing. H.L. and H.J.L. performed experiments on Mia PaCa-2 cells and mouse brain. H.L., N.T., and J.B.L. performed experiments on *E. coli* biofuel production. H.L., H.J.L., and J.X.C. wrote the manuscript with input from M.J.D., N.T., and J.B.L.

**Acknowledgement**: This work is supported by a DOE grant BER DE-SC0019387 to M.J.D., W.W. and J.X.C. and NIH grants R01GM118471, R33CA223581, R01AI141439 to J.X.C.

**Figure captions**

**Figure 1. Overview of fingerprint SRS microscopy by ultrafast tuning and spatial-spectral learning.** (a) The intrinsic cross-section of the coherent Raman scattering process and instrumentation define the conventional design space for SRS imaging, resulting in trade-offs between bandwidth (i.e., spectral resolution), speed and signal-to-noise ratio (SNR). Deep learning can expand the design space through computational methods, enabling high-speed, high-SNR fingerprint SRS imaging of living cells and large area tissues. (b) Setup of ultrafast delay-line tuning. A 55-kHz polygon scanner is used to scan the Stokes beam onto a Littrow-configured blazed grating to generate an SRS spectrum within 20 µs. By changing the angle between the grating blazed line and the laser scanned line, the effective delay range can be fine-tuned. PBS, polarizing beam splitter. QWP, quarter-wave plate. HWP, half-wave plate. PS, polygon scanner. DM, dichroic mirror. (c) Training a spatial-spectral residual net (SS-ResNet) deep neural network for SNR improvement, ground truth (GT) images are generated by averaging multiple acquisitions of the same field of view, equivalent to increasing the pixel dwell time. A trained network is then applied to recover the SNR of high-speed yet noisy images. (d) Spectral unmixing using least absolute shrinkage and selection operator (LASSO) to generate chemical maps.

**Figure 2. High-speed spectroscopic fingerprint SRS imaging of living Mia PaCa-2 cells.** (a) – (c) Fingerprint spectroscopic SRS imaging of fixed Mia PaCa-2 cells by raw single acquisition, SS-ResNet recovery from raw data and 100 images averaging (GT). (d) Zoom-in comparison of the same region of interest shown in white dashed boxes in (a)-(c). (e) Fingerprint SRS spectra of BSA, cholesterol and triglyceride serving as the pure spectral references of protein, cholesterol and unsaturated fatty acid. (f) Chemical maps of protein, cholesterol and fatty acid by pixel-wise LASSO unmixing from spectroscopic SRS images shown in (d). (g) Quantitative analysis of chemical mapping accuracy after network recovery. Box plots (n = 19) show the SSIM for raw vs. GT and network vs. GT of the three chemical channels. The boxes show interquartile range (IQR), the red line indicates medians, the black lines represent whiskers which extend to 1.5 times of the IQR and the red data points are the outliers exceeding the whiskers. (h) Fingerprint spectroscopic SRS imaging of living Mia PaCa-2 cells by SS-ResNet recovery of raw single acquisition and 100 averaging. (i-j) Cholesterol and fatty acid maps by LASSO unmixing from data in (g). Three significant motion artifacts are highlighted. (k-l) High-speed imaging of living Mia PaCa-2 cells in normal conditions after network recovery, followed by chemical maps of protein, cholesterol and fatty acid. (m-n) High-speed imaging of living Mia PaCa-2 cells with HPβCD treatment after network recovery, followed by chemical maps of protein, cholesterol and fatty acid. (o) Single-cell statistical analysis of the ratio between cholesterol and fatty acid over ~1000 cells in control and HPβCD treated group. Scale bars, 20 µm.

**Figure 3. Spectroscopic fingerprint SRS imaging of mouse whole brain.** (a) – (c) Fingerprint spectroscopic SRS imaging of mouse brain by raw single acquisition, network recovery and 100 averaging (GT). In each case, three representative images are shown. (d) Quantitative analysis of network recovery quality for mouse brain. Box plots (n = 15) show the NRMSE and SSIM for raw vs. GT and network vs. GT. (d)-(f) Protein, fatty acid and cholesterol maps of a mouse whole brain slice after SS-ResNet recovery. (g) Mouse whole brain composite chemical maps consisting of protein (blue), fatty acid (red) and cholesterol (green). Different colors indicate different percentage concentrations from the three channels. (h) Zoom-in images of different mouse brain areas. Circled regions in the DG area include rare cells with the high cholesterol content. LH, lateral hypothalamus; BM, basal amygdaloid; VP, ventral posterior nucleus; CPu, caudate putamen; DG, dentate gyrus; HC, hippocampus.

**Figure 4. High-speed imaging of *E. coli* biofuel production strains.** (a) Fingerprint SRS spectra of BSA, pure limonene and pinene samples. (b)-(d) Fingerprint spectroscopic SRS imaging of *E. coli* by raw single acquisition, SS-ResNet recovery from raw data and 50 averaging (GT). In each case, three representative images are shown. (e) Quantitative analysis of network recovery quality for the *E. coli* dataset. Box plots (n = 13) show the NRMSE and SSIM for raw vs. GT and network vs. GT. network recovered image at 1650 cm$^{-1}$ and chemical mappings of protein, limonene and pinene for (f) wild type, (g) limonene production and (h) pinene production strains. Scale bars, 10 µm.

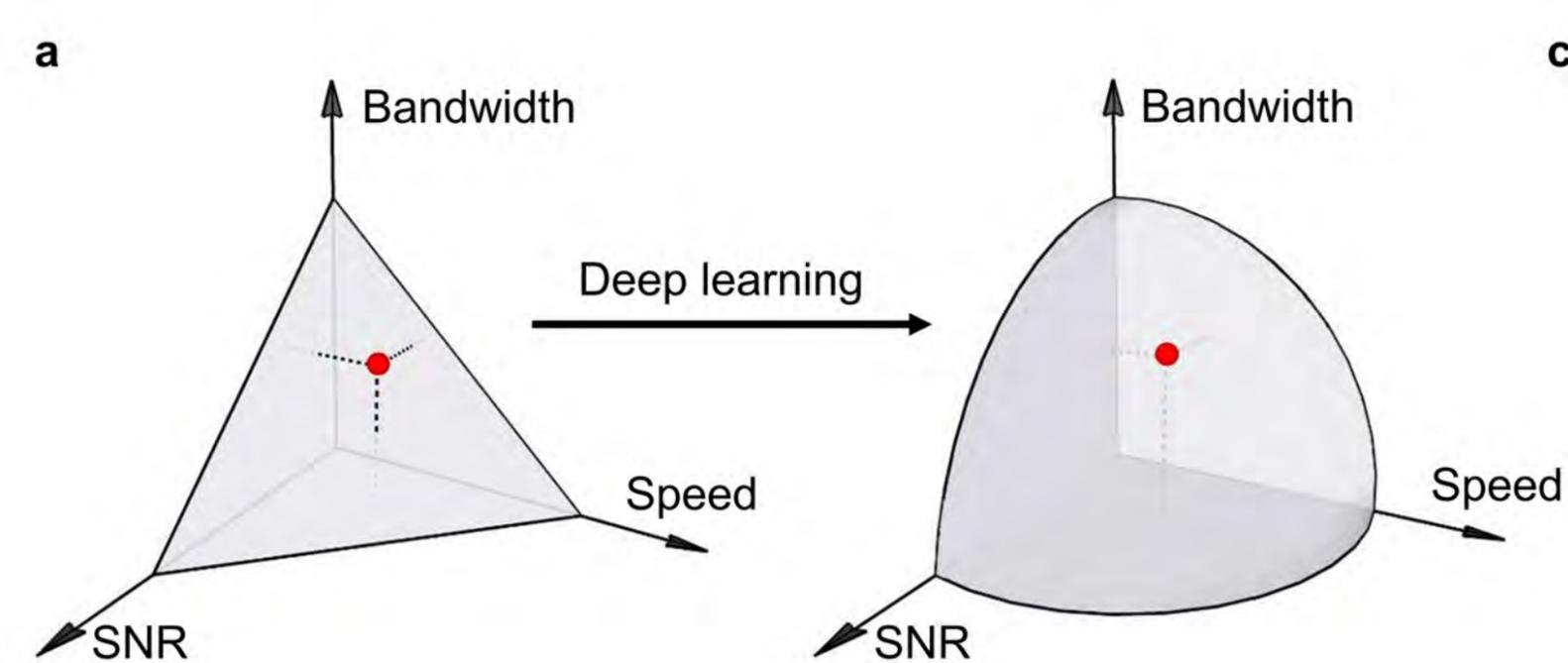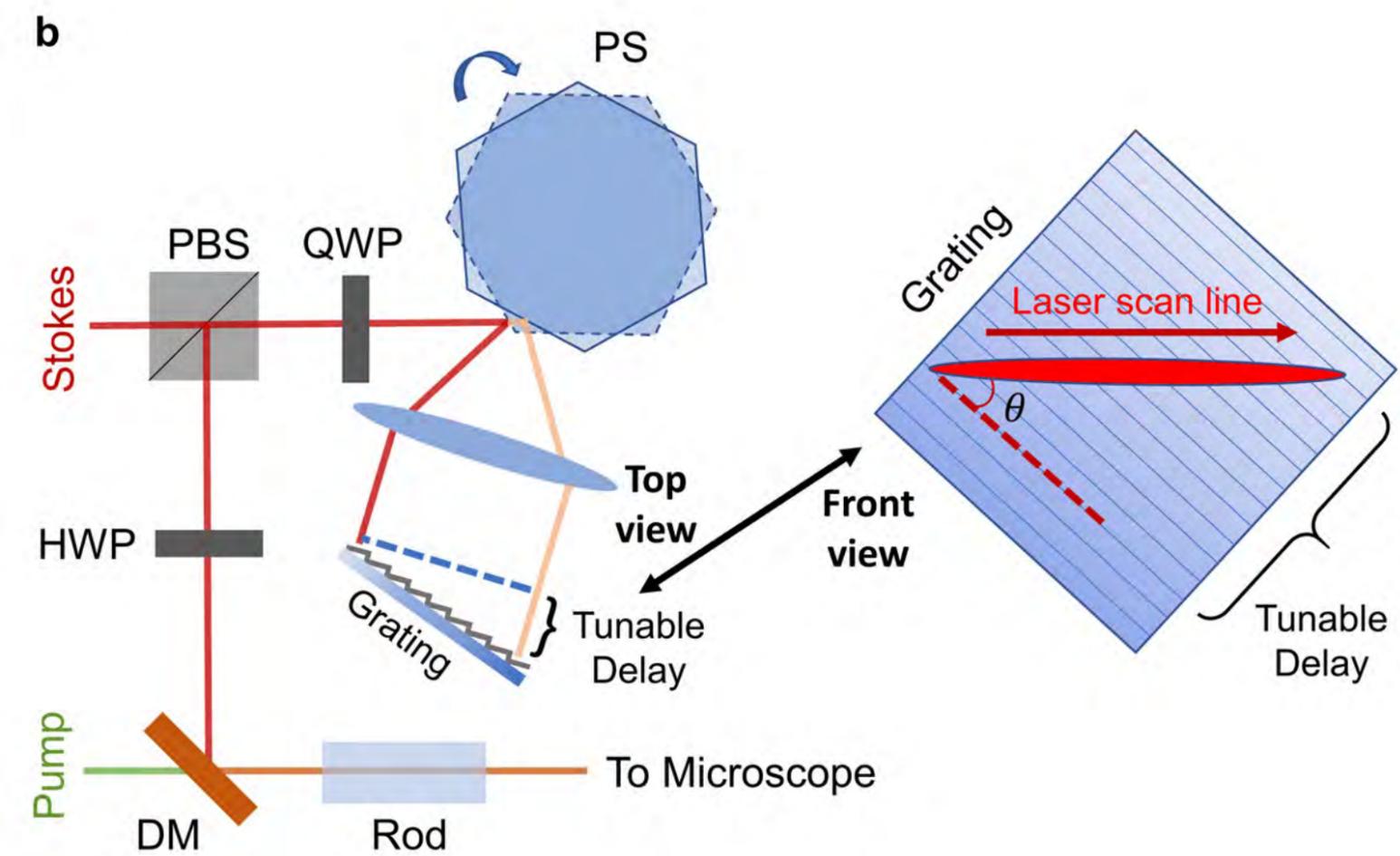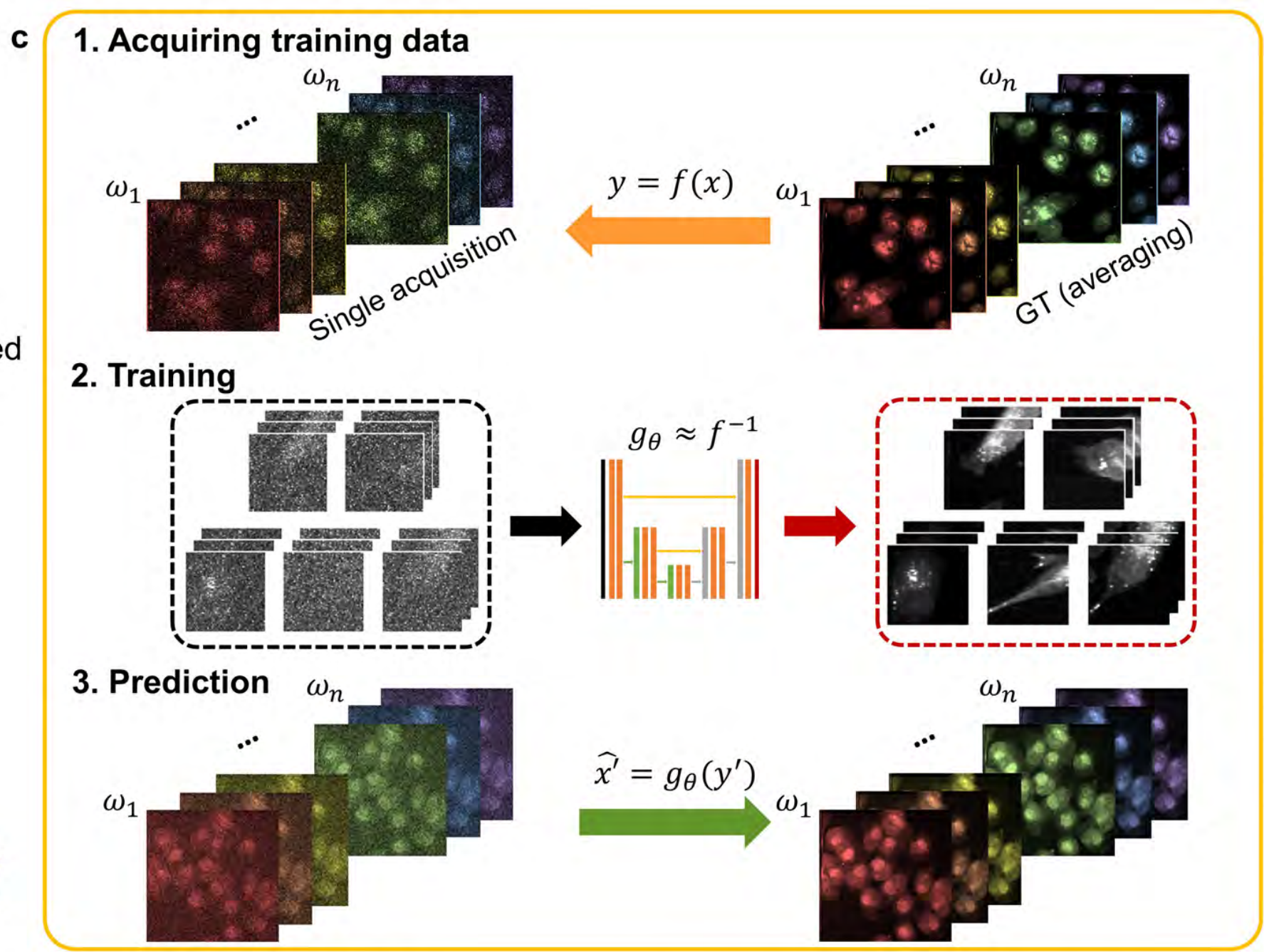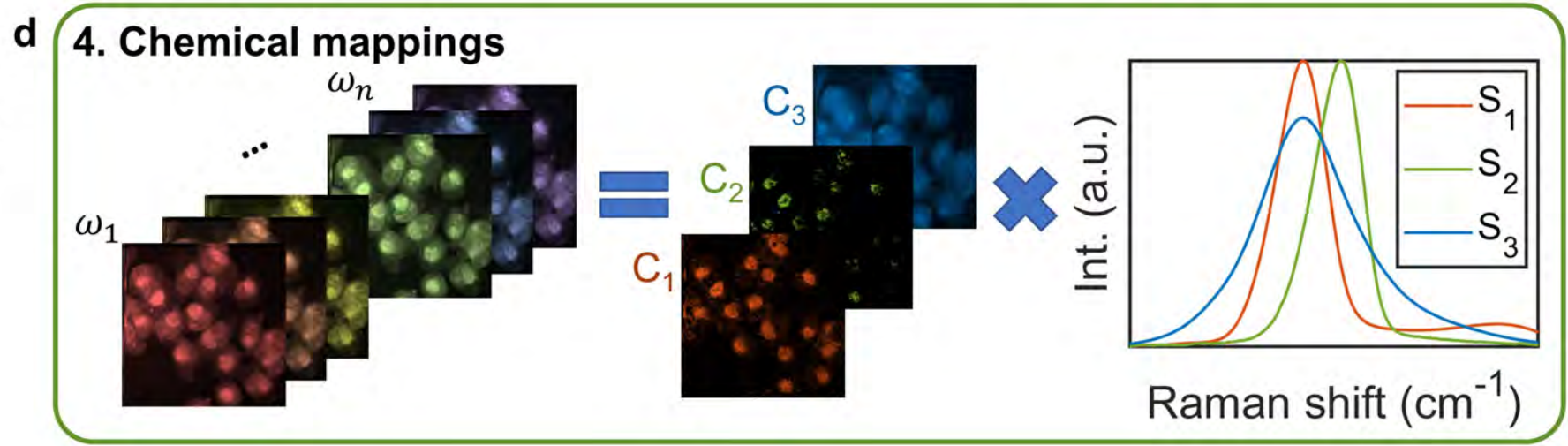

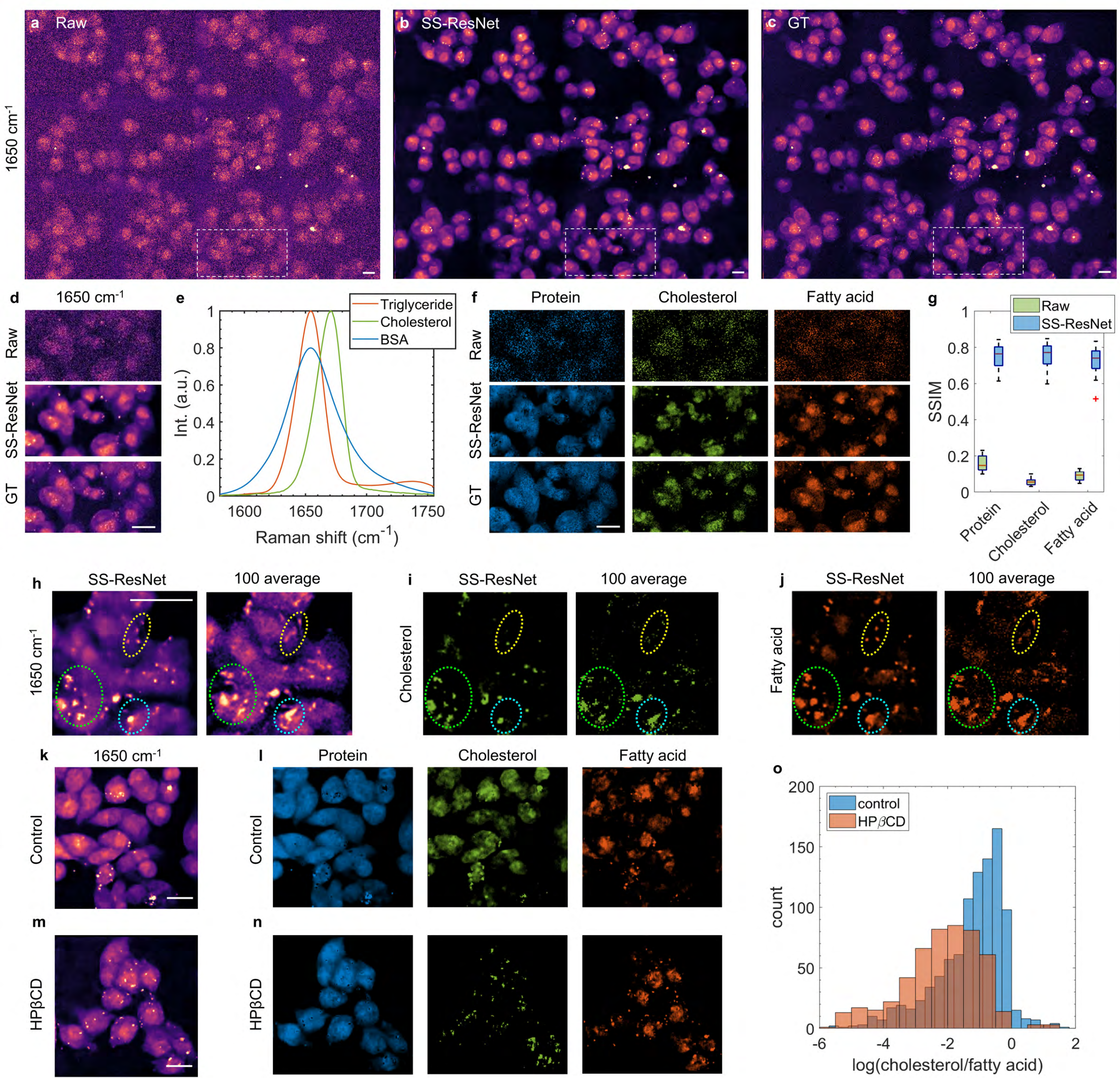

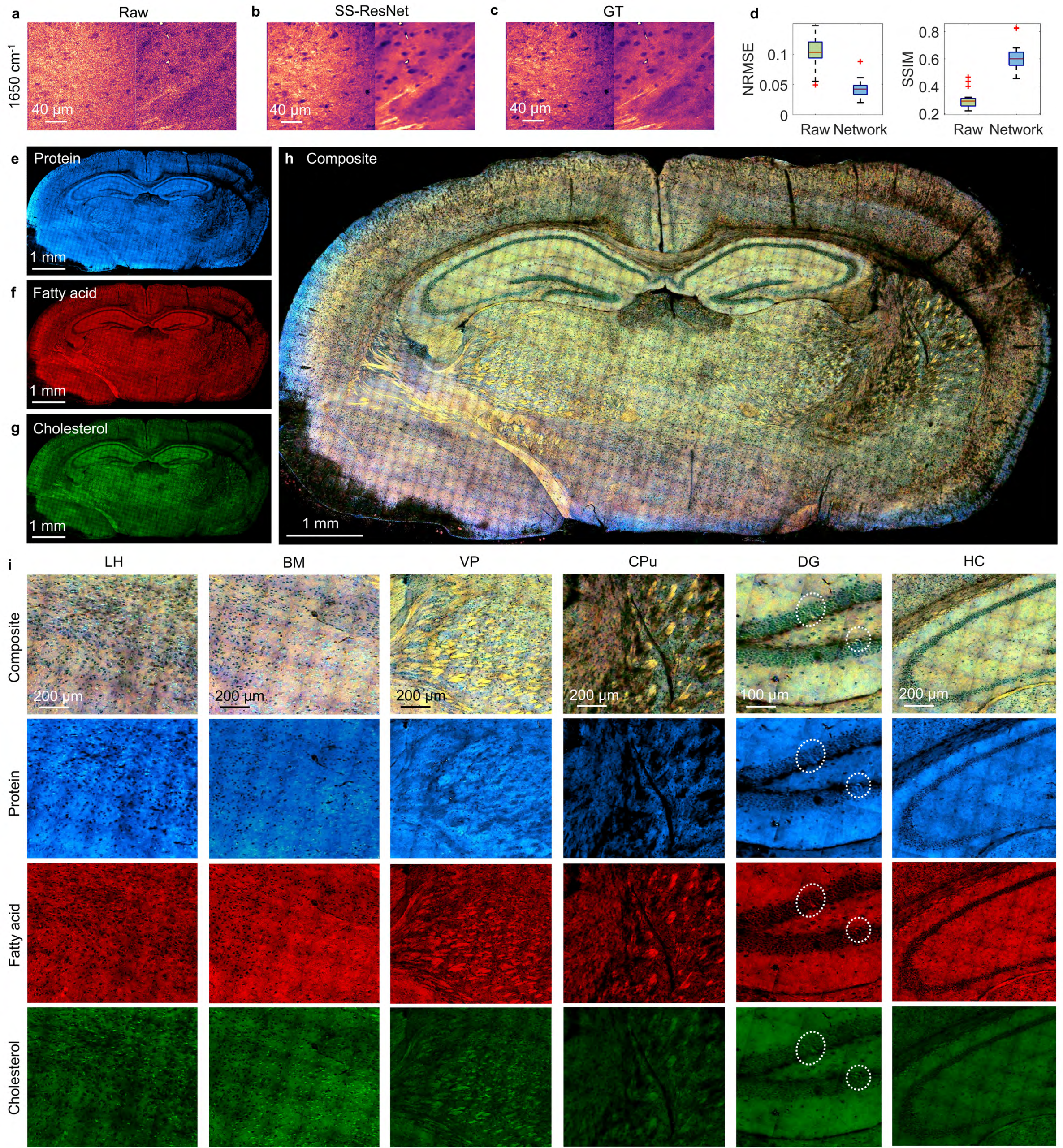

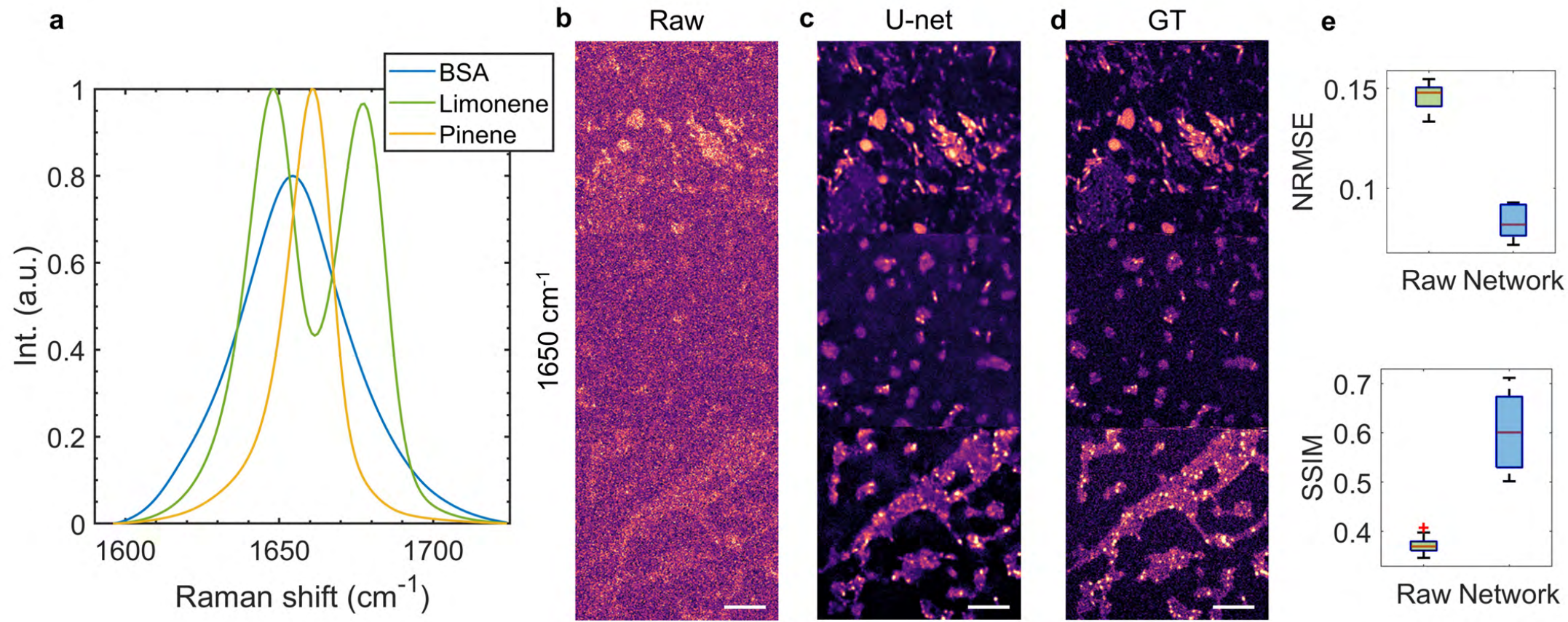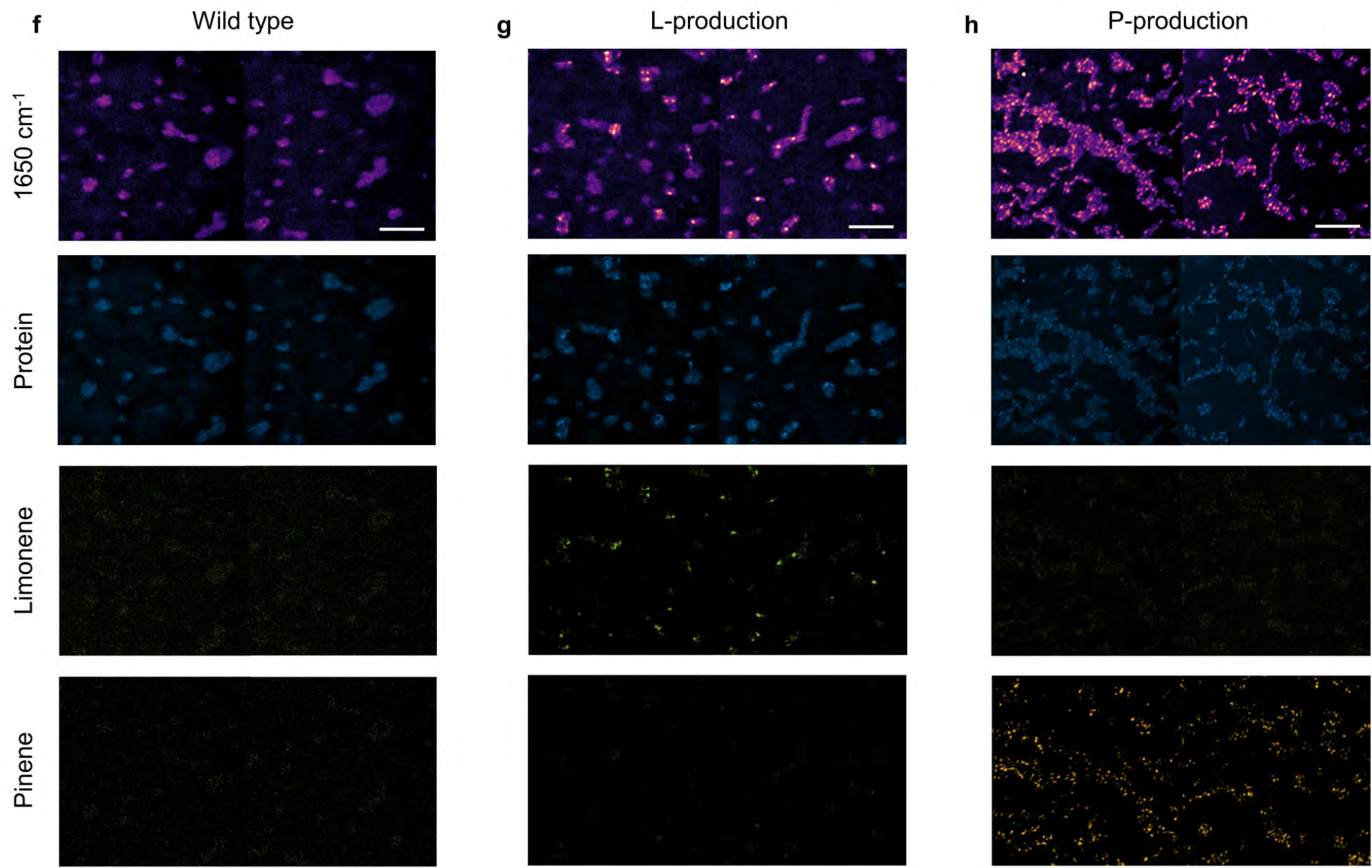

# Fingerprint Spectroscopic SRS Imaging of Single Living Cells and Whole Brain by Ultrafast Tuning and Spatial-Spectral Learning

Haonan Lin et al.

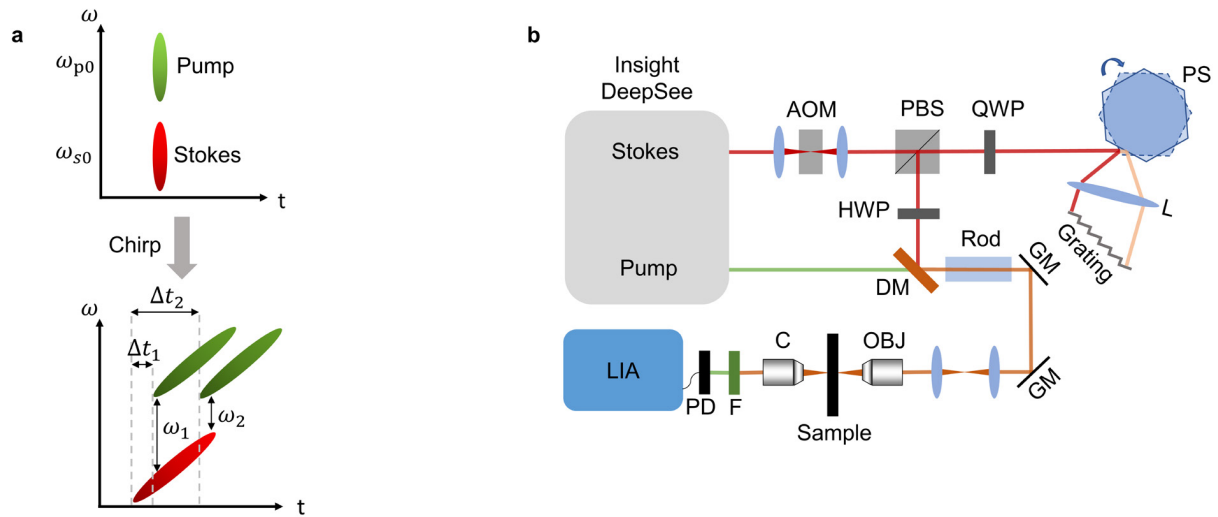

**Supplementary Figure 1. Optical setup of ultrafast tuning SRS system**. (a) Concept of spectral focusing. (b) Optical setup. AOM, acousto-optic modulator; C, condenser; F, filter; GM, galvo mirro; HWP, half-wave plate; L, lens; LIA, lock-in amplifier; OBJ, objective; PBS, polarizing beam splitter; PD, photodiode; PS, polygon scanner; QWP, quarter-wave plate.

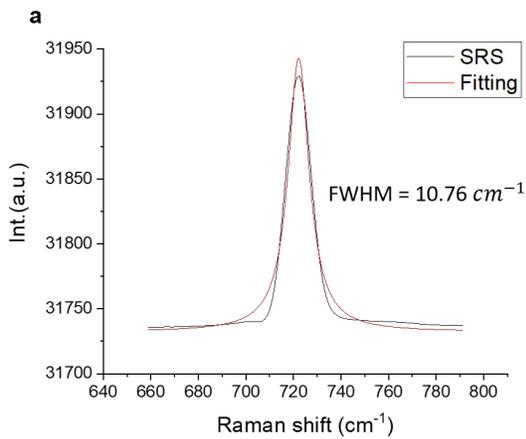
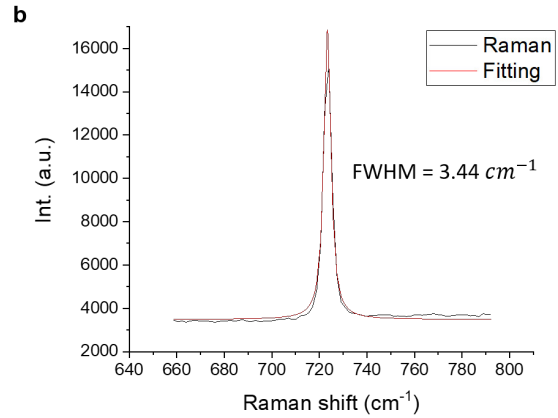
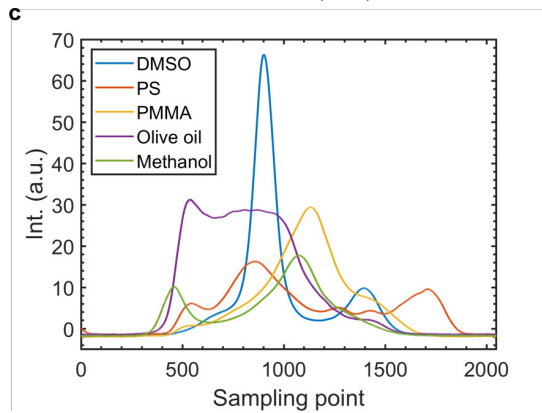
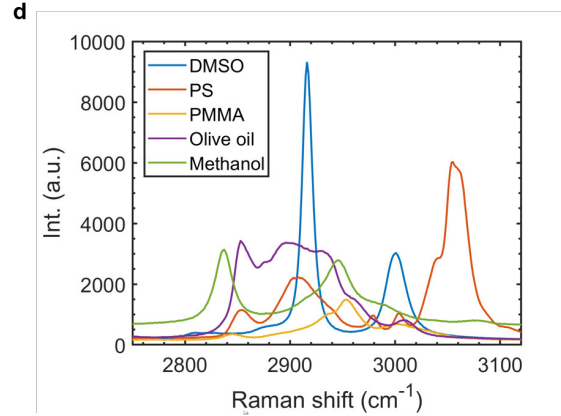
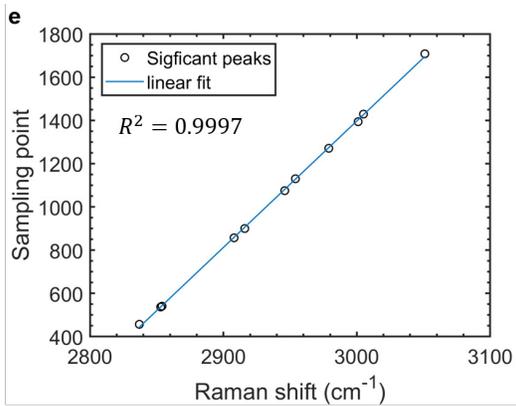
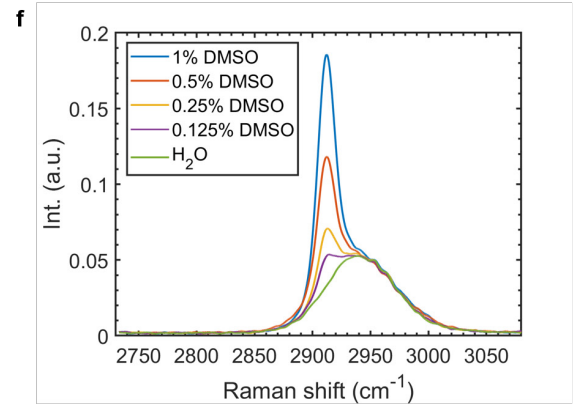
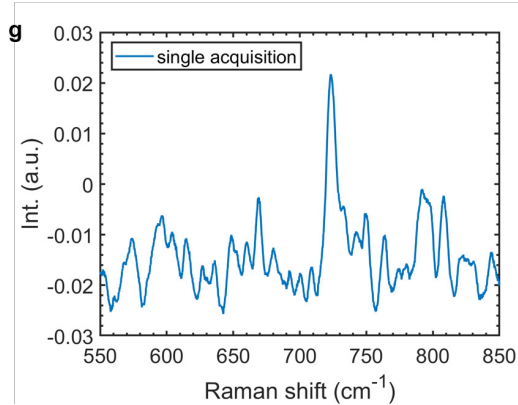
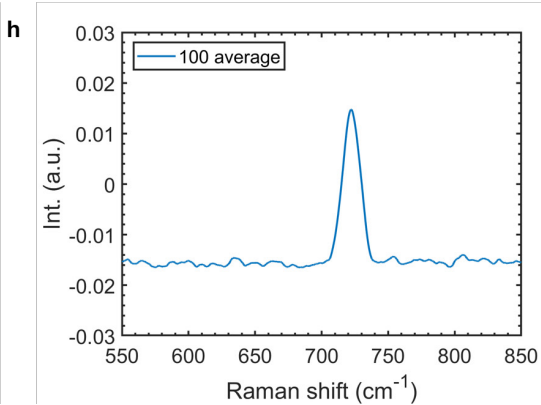

**Supplementary Figure 2. Spectral resolution, spectral linearity and sensitivity of ultrafast delay-line tuning SRS.** (a) SRS spectrum of adenine and Lorenz one-peak fitting. (b) Spontaneous Raman spectrum of adenine and Lorenz one-peak fitting. (c) SRS spectra of DMSO, PS, PMMA, olive oil and methanol obtained by polygon delay-line scanning system. (d) Spontaneous Raman spectra of the same chemicals. (e) Mapping of Raman shifts to sampling point number of the digitizer (corresponding to acquisition time from sampling trigger). (f) SRS spectra of water and DMSO solutions with different concentrations. (g) Fingerprint SRS spectra of adenine by single acquisition (h) Fingerprint SRS spectra of adenine by 100 averaging.

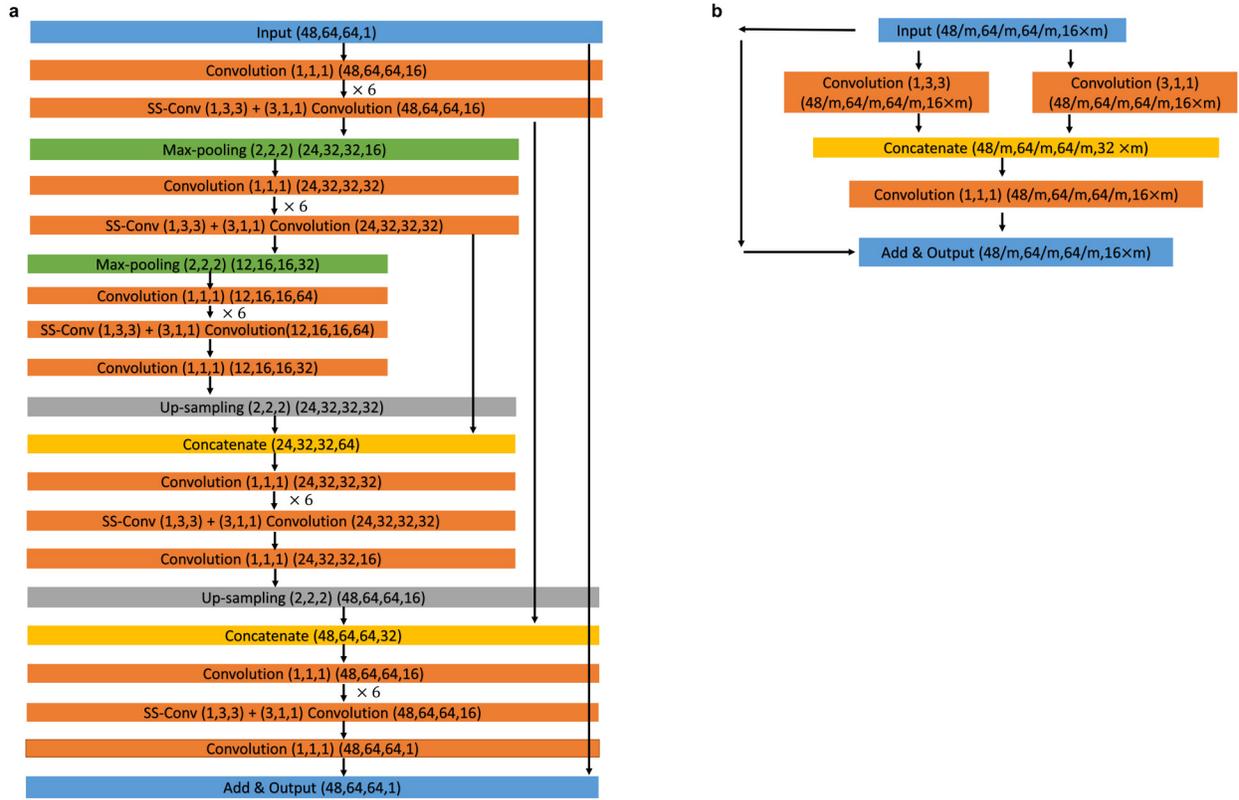

**Supplementary Figure 3**. **SS-ResNet structure for SNR recovery of spectroscopic images**. (a) Complete structure of the network. Orange layers represent either conventional or spatial-spectral convolution. Green layers represent max-pooling layers with (2,2,2) kernel size. Gray layers stand for (2,2,2) up-sampling while the yellow layers concatenate up-sampled feature maps with corresponding feature maps in the previous encoder level. (b) Structure of each SS-Conv layer. The input is separately filtered spatially and spectrally through a (1,3,3) and (3,1,1) convolution. The two outputs are concatenated and passed through a (1,1,1) convolution to reduce the feature channels. Adding with the input yields the final output of the convolution layer.

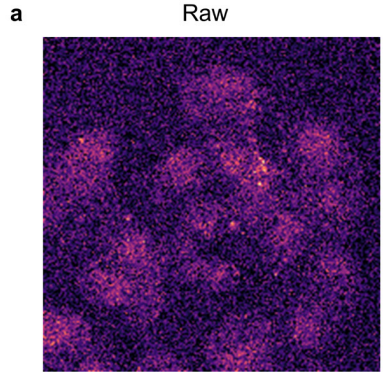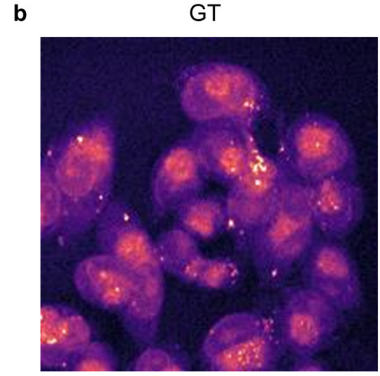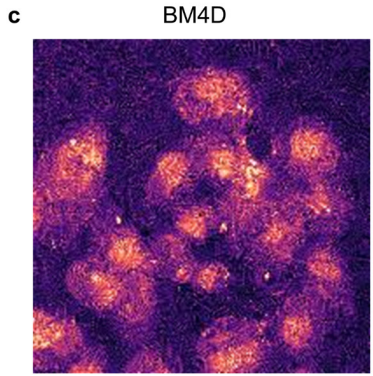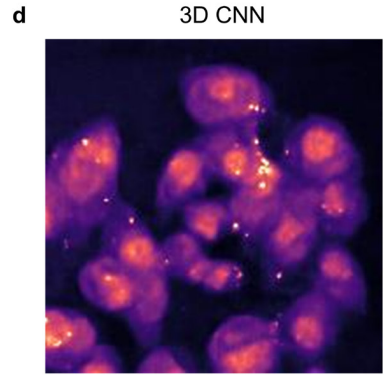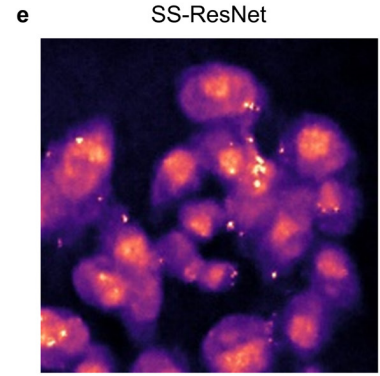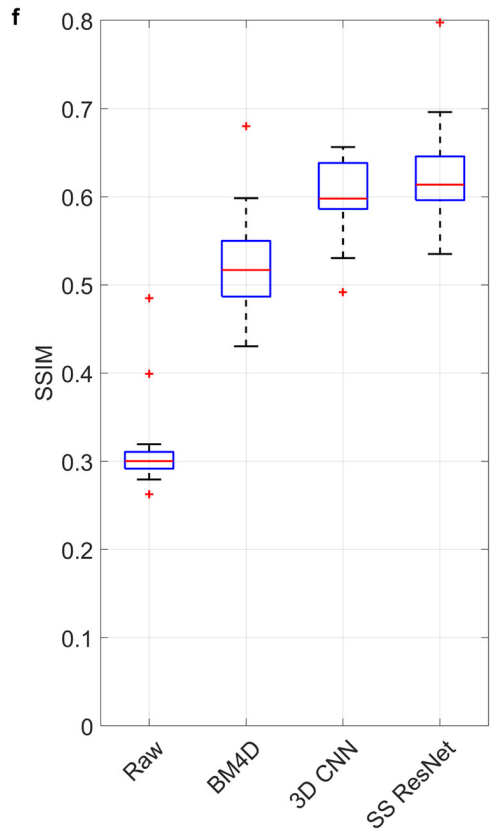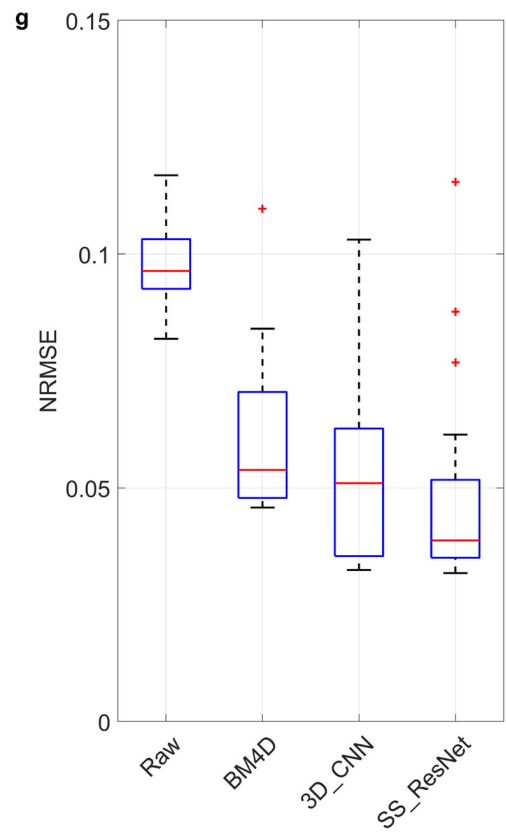

**Supplementary Figure 4. Comparing SS-ResNet with 3D CNN and BM4D**. (a) Raw single acquisition data. (b) High-SNR GT data by averaging the same FOV 100 times. (c) BM4D denoising from raw image. (d) 3D CNN recovery from raw image. (e) SS-ResNet recovery from raw image. A comparison between BM4D, CNN and SS-ResNet demonstrates the advantageous performance of SS-ResNet recovery, which is free of denoising-induced artifacts that distort the cell structures while maintaining most of the fine structures. (e) Quantitation of prediction error for fixed Mia PaCa-2 validation dataset, including 19 spectroscopic images of size 200×200×128. Box plots show RMSE (lower is better) and SSIM (higher is better) for the raw, BM4D, 3D CNN and SS-ResNet recovery. The boxes show interquartile range (IQR), the red line indicate medians, the black lines represent whiskers which extend to 1.5 times of the IQR, the red datapoints are outliers exceeding the whiskers.

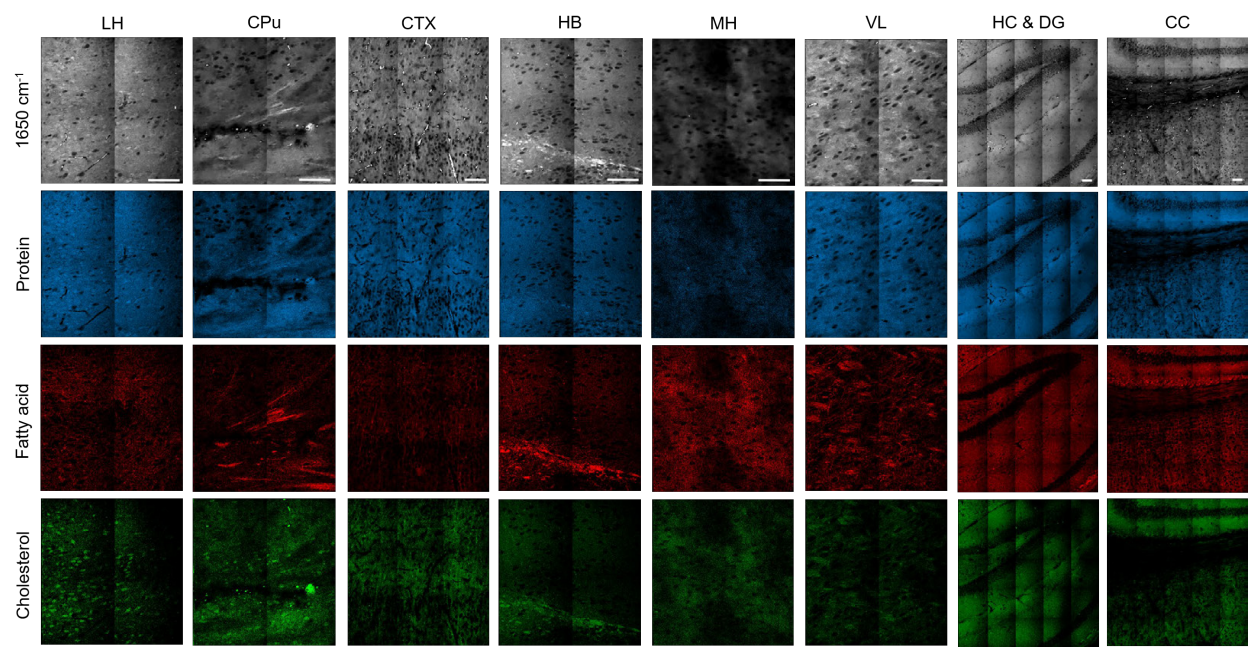

**Supplementary Figure 5**. **Whole brain training set**. GT (100 average) at 1650 cm$^{-1}$ and downstream chemical maps by LASSO unmixing. Scale bars, 50 µm.

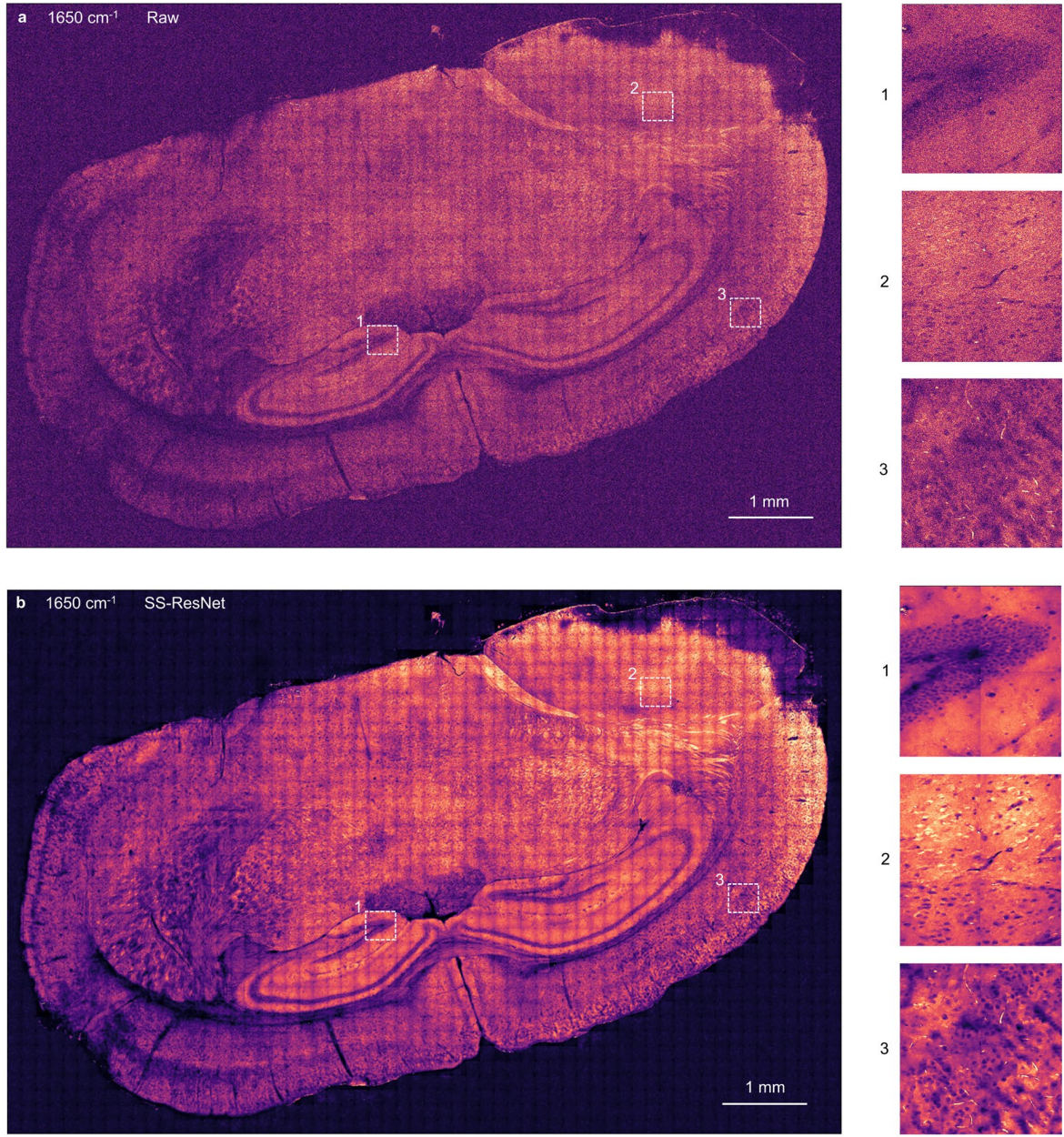

**Supplementary Figure 6**. **Mouse whole brain image at 1650 cm$^{-1}$**. (a) Raw image. (b) Network recovery from raw image.

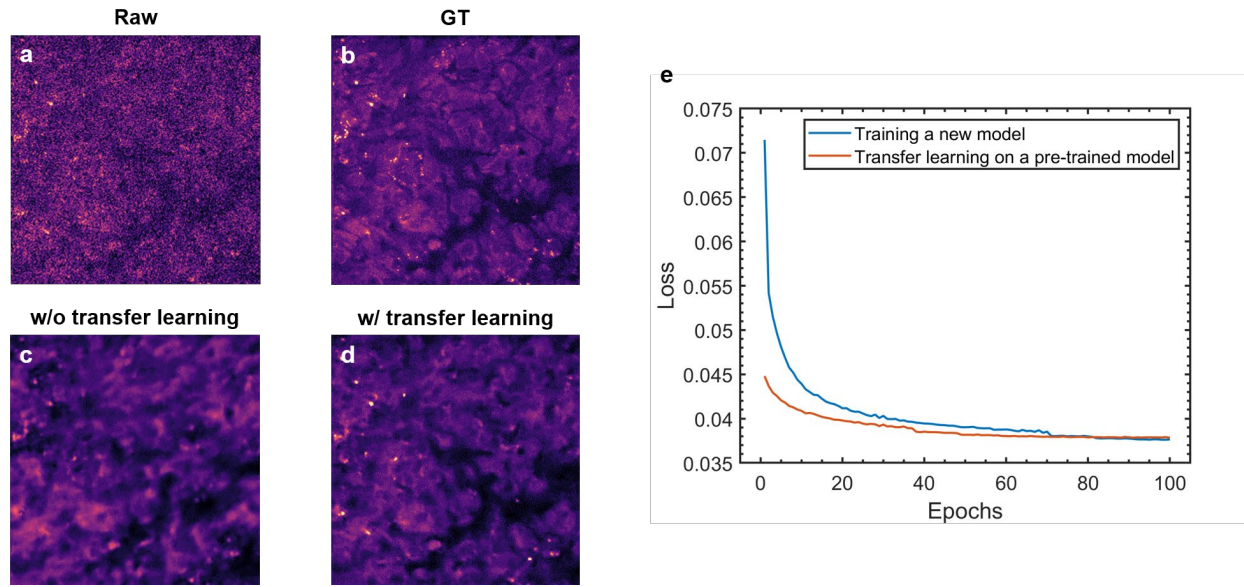

**Supplementary Figure 7**. **Transfer learning.** (a) Raw prostate cancer image (b) GT image by averaging 100 times. (c) Recovered prostate tissue image by applying a network trained on Mia PaCa-2 cells images. (d) Recovered prostate tissue image after transfer learning.

**Video Captions**

**Supplementary video 1**. Video of live Mia PaCa-2 cells at 1650 cm$^{-1}$ by raw acquisition.

**Supplementary video 2**. Video of live Mia PaCa-2 cells at 1650 cm$^{-1}$ by network recovery from raw acquisition.

**Supplementary video 3.** Video of protein chemical maps of live Mia PaCa-2 cells.

**Supplementary video 4.** Video of fatty acid chemical maps of live Mia PaCa-2 cells.

**Supplementary video 5.** Video of cholesterol chemical maps of live Mia PaCa-2 cells.